\documentclass[doublecol]{epl2}

\usepackage{graphicx}
\usepackage{dcolumn}
\usepackage{bm}
\usepackage{hyperref} 
\usepackage{amsmath}
\usepackage{amsfonts}     
\usepackage{graphics}  
\usepackage{psfrag} 
\usepackage{graphicx}    
\usepackage{epsfig}    
\usepackage{rotating}  
\usepackage[latin1]{inputenc}
\usepackage{color}
\usepackage{rotating}
  
\title{Current reversal and exclusion processes with history-dependent random walks}

\author{Johannes H.P. Schulz\inst{1} \and Anatoly B. Kolomeisky\inst{2} \and Erwin Frey\inst{1}}
\shortauthor{Johannes H. P. Schulz \etal}

\institute{
  \inst{1} Arnold Sommerfeld Center for Theoretical Physics and Center for NanoScience, Department of Physics, Ludwig-Maximilians-Universit\"at M\"unchen, Theresienstra{\ss}e 37, D-80333 M\"unchen, Germany\\
  \inst{2} Department of Chemistry, Rice University, Houston, Texas 77005, USA
}

\date{\today}
 
\abstract{
A class of exclusion processes in which particles perform history-dependent random walks is introduced, stimulated by dynamic phenomena in some biological and artificial systems. The particles locally interact with the underlying substrate by breaking and reforming lattice bonds. We determine the steady-state current on a ring, and find current-reversal as a function of particle density. This phenomenon is attributed to the non-local interaction between the walkers through their trails, which originates from strong correlations between the dynamics of the particles and the lattice. We rationalize our findings within an effective description in terms of quasi-particles which we call front barriers. Our analytical results are complemented by stochastic simulations.
}

\pacs{05.40.-a}{Fluctuation phenomena, random processes, noise, and Brownian motion}
\pacs{02.50.-r}{Stochastic processes}
\pacs{87.10.Mn}{Stochastic modeling}

\begin{document}
\maketitle

\section{Introduction}

Brownian motion can be rectified in manifold ways \cite{Julicher:1997p28955, reimann-2002-361, Haenggi:2009p29024, Kay:2007p28153}. One design principle exploits interactions between Brownian particles and a substrate leading to local changes in the substrate properties which in turn affect the particle's motion. For example, the processive uni-directional motion of collagenase on collagen fibrils \cite{Saffarian:2004p28152, Saffarian:2006p28147} has been argued to be due to a \emph{burnt-bridge mechanism} \cite{Mai:2001p27900}: collagenase, as it diffuses along collagen fibrils, cleaves its track at some recognition sites so that it always ends up behind the cleavage site. This then acts as a burnt bridge biasing collagenase motion. Similarly, motivated by the translocation of Holliday junctions, asymmetric nucleation of hydrolysis waves has been proposed as a track-mediated mechanism driving directed motion  \cite{Klapstein:2000, Lakhanpal:2007}. Recently, artificial systems employing bipedal DNA motors have been engineered using a track-mediated rectification principle \cite{Omabegho:2009p67}. Moreover, synthetic molecular systems, termed molecular spiders \cite{Pei:2006p27865, Samii:2010p021106, Semenov:2011p021117}, have been constructed utilizing a similar mechanism. They consist of an inert body and catalytic legs. The legs are DNA enzymes which, upon binding to a complementary DNA substrate, cleave it into two shorter products that then have a lower affinity for the legs. The spider's motion thus changes the biochemical properties of the molecular track, and its trail consists of sites with enhanced dissociation rates. When interacting with precisely defined environments such molecular spiders even perform some elementary robotic behavior \cite{Lund:2010p28160}. 

Both of these examples are \emph{non-Markovian stochastic processes}: through cleaving the track or changing kinetic parameters of the substrate the system acquires a memory of the path traced out by the particle. Such processes belong to a general class of random walks on lattices with weighted bonds or sites. Well-studied models include the self-avoiding random walk \cite{Madras_Slade:book}, the reinforced random walk \cite{DAVIS:1990p28173}, and the excited random walk \cite{Antal:2005p28234}. These random walks with memory show unusual behavior, like anomalous diffusion or spatial confinement.

We are interested in \emph{collective phenomena} emerging from the interaction of many non-Markovian random walkers. Such systems are genuinely distinct from their Markovian counterparts like, e.g., the symmetric or asymmetric exclusion processes (ASEP) \cite{Derrida:1998p29316} or Brownian ratchets \cite{DERENYI:1995p30245}. Whereas in the latter case interactions are local in space and time, they are non-local for non-Markovian random walkers because they mutually influence each other through their trails, i.e., changes induced in the track while moving over it. Since this implies strong correlations between the dynamics of the particles and the lattice, we expect novel behavior as compared to Markovian systems where inhomogeneities are either quenched, i.e., static on the time scale of particle dynamics, or annealed, i.e., fast and decoupled from the particle dynamics. We focus on the asymptotic dynamics at long times, where a stationary state is reached in both the particle and lattice dynamics. The main quantity of interest is the current as a function of the model parameters such as the kinetic rates and the particle density. We find a particularly interesting phenomenon: a current reversal with respect to the particle concentration.

\section{The model}
\begin{figure}[htp]
\centering 
\includegraphics[width=\columnwidth]{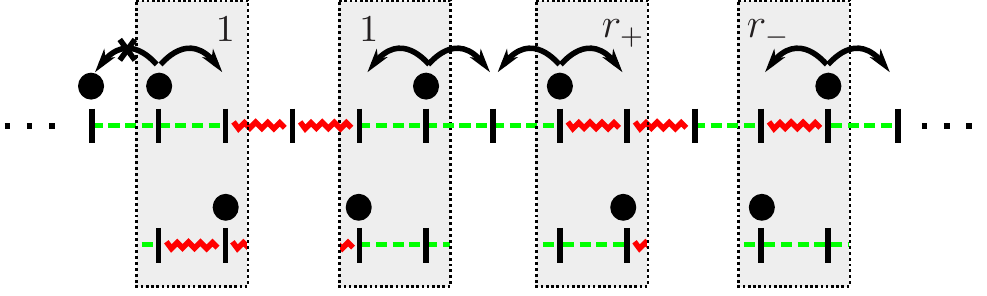} 
\caption{\label{fig:model} \emph{Illustration of the burnt-bridge exclusion process.} The one-dimensional lattice consists of bonds, which can be either in an intact (dashed green) or in a burnt state (zig-zag red). Only upon crossing an intact bridge to the right it gets burnt. Passing a burnt bridge to the right or left repairs a bridge at rate $r_\pm$. Particles are subject to on-site exclusion.}
\end{figure}

In this work, we generalize the reversible  \emph{``Burnt-Bridge model''} (BBM) \cite{Mai:2001p27900, Antal:2005p27853, Morozov:2007p9187, Artyomov:2008p27887, Artyomov:2010p27938} to a non-Markovian many body system where particles interact through their trails and on-site exclusion. In analogy to the asymmetric simple exclusion process (ASEP) we term the model \emph{burnt-bridge exclusion process} (BBEP). In this model a fixed number $N$ of random walkers moves along a \emph{periodic} one-dimensional lattice with sites $i = 1, \ldots ,L$ connected by bonds which can be in either of two states: intact or broken; see fig.~\ref{fig:model}.  
Each site is either occupied by a single particle or is empty. Bonds are subject to local interaction with crossing particles which may ``destroy'' a formerly intact bond, leaving it in a broken state (``burnt bridge") or ``heal" a burnt bridge restoring it into an intact state again. If a bond is intact a particle may cross it with rate $1$ in either direction (fixing the time scale). We assume a totally asymmetric burning mechanism\footnote{This assumption is made for simplicity. All main conclusions of our analysis remain valid for more complicated versions of the model \cite{LongPaper}.}: A particle will \emph{always} burn an intact bond when crossing it in the positive direction (to the right), while  it will \emph{never} damage an intact bond, when crossing it in the negative direction (to the left). Once a bridge has been burnt, it poses an obstacle for any approaching particle. However, these are only transient obstacles for the particle traffic: We assume that broken bonds are ``repaired'' at reduced rates \(r_\pm \leq1\), that depend on the crossing direction as explained in fig.~\ref{fig:model}. For the parameter regime $r_- \gg r_+$ the asymmetric repair mechanism and the burnt-bridge mechanism steer the particles in opposite directions. In this case, the non-Markovian character of the random walk may show its most explicit manifestation as a current reversal, both with respect to the lattice length and the number of particles.

\section{The front barrier picture}
\begin{figure}
 \includegraphics[width=\columnwidth]{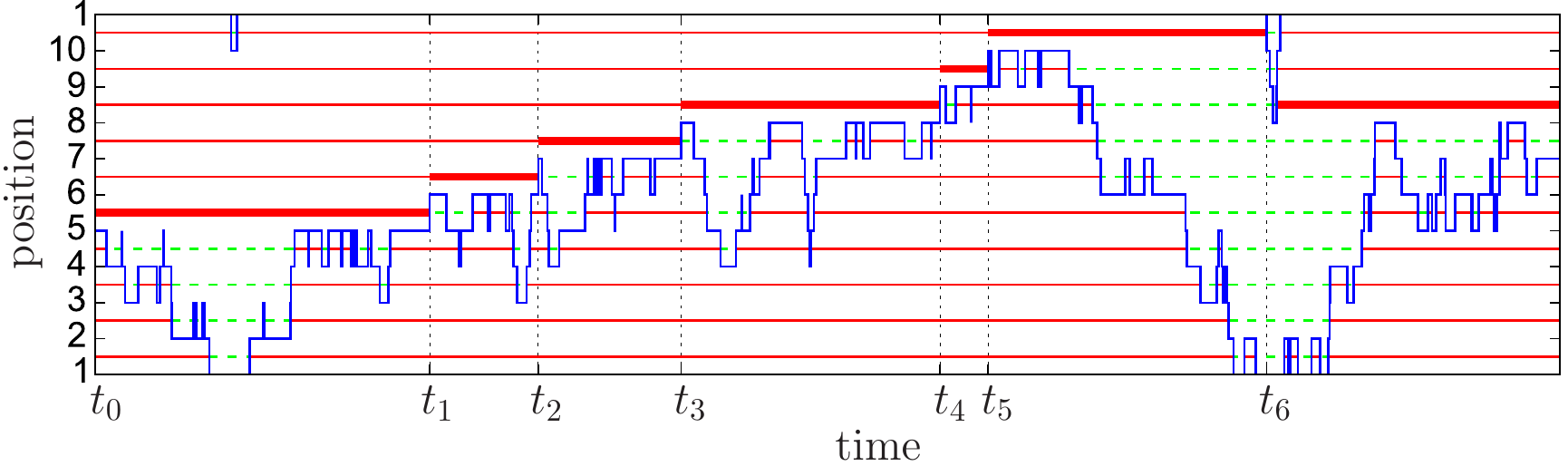}
 \caption{\label{fig:trajectory_single_walker} Typical trajectory of a random walker (blue) with repair rates \(r_-=0.65\), \(r_+=0.15\), and lattice size \(L=10\). Horizontal lines indicate the state of the bonds: intact (dashed green) or burnt (solid red). The ``front barrier" is marked by a thick red line. \emph{Pushing events} occur at times \(t_0\),...,\(t_5\),  and two \emph{pulling events} happen right after \(t=t_6\).}
\end{figure}

Already the dynamics of a single walker pose a highly difficult mathematical problem which can be solved exactly only in some special cases \cite{Artyomov:2010p27938}. Here we show that for the parameter regime, \(r_- \gg r_+\), where the asymmetric repair mechanism and the burnt-bridge mechanism are antagonists, there is an effective description in terms of a kind of quasi-particle which we call the ``front barrier". To develop this picture we start our analysis by exploring the characteristic features of a random walker's  trajectory; see fig.~\ref{fig:trajectory_single_walker}.
We start the particle's journey on a lattice with all bonds broken such that it finds itself trapped between two partially reflecting barriers.  Since \(r_- \gg r_+\) broken bonds are more ``transparent" in the backward direction, the walker most probably bounces off the broken bond in the forward direction, and repairs the bond in the backward direction, c.f. fig.~\ref{fig:trajectory_single_walker}.  We call the strongly reflecting bridge in the forward direction \emph{``front barrier"} (FB). In every step backward bridges are repaired, while they are burnt for each forward step. Thus, the particle finds itself not only initially, but for some extended time period in front of a burnt bridge and behind the FB; here  during \([t_0,t_1]\). Since $r_- < 1$ the random walker is biased to move towards the FB but bounces off several times before it finally happens to repair it at time $t_1$. Now, it finds itself in a similar situation as at time $t_0$, i.e., behind a strongly reflecting barrier. With an intact bridge behind, it most probably again bounces off the reflective wall. When doing so, the particle does not burn the intact bridge, as this is a backward crossing event. Up to this minor detail the particle dynamics during the time interval \([t_1,t_2]\) follows the same principles as for \([t_0,t_1]\). The FB has now been \emph{pushed forward} by the walker repairing it at \(t_1\).

This pattern of movement could basically repeat itself indefinitely; see time intervals \([t_0,t_1]\), \([t_1,t_2]\),..., \([t_4,t_5]\) in fig.~\ref{fig:trajectory_single_walker}. However, since the lattice is finite, the random walker may at some point in time also travel the whole length of the ring in the backward direction, and eventually arrive on the \emph{weakly} reflecting side of the FB. From there, it can easily pierce (and thereby remove) the FB. At this time, \(t=t_6\), there are no more burnt bridges left on the lattice, and the random walk becomes symmetric, but only until crossing of a bridge in the forward direction induces this bridge to burn. In our sample trajectory the random walker makes two backward steps before it turns forward again. It leaves behind a trace of burnt bridges and eventually arrives at a newly formed, strongly reflecting FB, which has been \emph{pulled back} by two steps.
\begin{figure}
\centering
\includegraphics[width=\columnwidth]{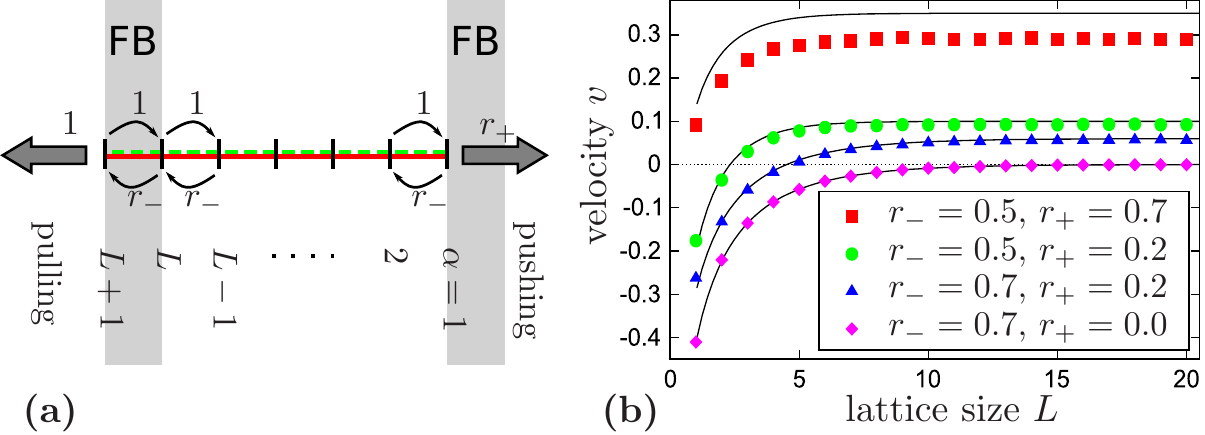}
 \caption{\label{fig:manyb_singlew} (a) \emph{Front barrier picture} as explained in the text.
(b) The velocity \(v\) of a single random walker as a function of lattice length $L$. Simulation results (symbols) are given for different sets of repair rates. Solid lines are obtained from the front barrier picture, eq.~\eqref{eq:velocity_single_walker}.}
\end{figure}
In summary, the dynamics of the FB is determined by the antagonism between pushing and pulling events. This leads to a current reversal as a function of the lattice length $L$; cf. fig.~\ref{fig:manyb_singlew}. For long rings, \(L \gg 1\), the particle, due to the weak bias in the positive direction, is localized in the immediate vicinity of the front barrier. Then, the dynamics are dominated by pushing events, and the velocity is positive. In contrast, shortening the lattice increases the relative frequency of pulling events. The probability of such events heavily depends on the random walker's ability to diffuse over the whole length of the lattice. Therefore, for small enough lattices, pulling events may at some point outnumber pushing events, resulting in a negative velocity.

These heuristic observations can also be formulated quantitatively. For the parameter regime $r_+ \ll r_- < 1$ the state space most relevant for the dynamics is highly reduced. Instead of all the \(2^{L}\) possible bridge configurations, only ``block configurations'' are statistically significant. These consist of two continuous regions on the lattice where the bridges are either burnt or intact, a ``burnt block'' and an ``intact block'', separated by a front barrier.  The main reason for this reduction is a separation of time scales: since repair of bridges in the forward direction are rare events ($r_+ \ll r_- < 1$) a random walker finds itself most of the time behind a strongly reflecting bridge, the FB\footnote{For large \(r_+\), the approximation fails since it becomes more and more likely that a random walker pushes the FB forward by not only one but many steps. Then, after reflection at the FB, it is able to constitute a \emph{new} block of burnt bridges \emph{inside} the previously intact block, and the picture with a single block configuration breaks down.}. In this FB picture the random walker's effective states are $\alpha = 1, \ldots, L$, where it is located at a distance $\alpha$ behind the FB, and $\alpha=L+1$, where it has crossed the FB from the right and finds itself on a fully intact lattice; see fig.~\ref{fig:trajectory_single_walker} and fig.~\ref{fig:manyb_singlew}a. In the latter state the FB is \emph{pulled backward} at rate $1$, while in state $\alpha=1$ the FB is \emph{pushed forward} at rate $r_+$. Hence the velocity of the FB and with it the average particle velocity is given by
\begin{equation}
\label{eq:velocity_1rw}
v = r_+ \cdot m_1 - 1 \cdot m_{L+1} \, ,
\end{equation}
where $m_\alpha$ denotes the probability to find the walker in state $\alpha$. Since the boundary states $\alpha=1, \, L+1$ are totally reflecting, the probability current between two states is strictly zero: $j_\alpha =1 \cdot m_{\alpha+1}-r_- \cdot m_\alpha = 0$. This implies that $m_\alpha$ decays exponentially, $m_\alpha =\left(r_-\right)^{\alpha-1} m_1$. With the normalization of the probability distribution, $\sum^{L+1}_{\alpha=1} m_\alpha=1$, this gives $ m_1 = {(1-r_-)}/{(1-r_-^{L+1})}$ 
such that the average particle velocity, eq.~\eqref{eq:velocity_1rw}, reads
\begin{equation} 
\label{eq:velocity_single_walker}
v (L) = \frac{1-r_-}{1 - r_-^{L+1}}\left[r_+ - r_-^{L}\right] \, .
\end{equation}
These results can be generalized to lattices containing both strong and weak links, where \emph{strong links} are unaffected by the random walker and \emph{weak links} correspond to the bridges considered here \cite{LongPaper}.

\begin{figure}[hbt]
\centering
\includegraphics[width=\columnwidth]{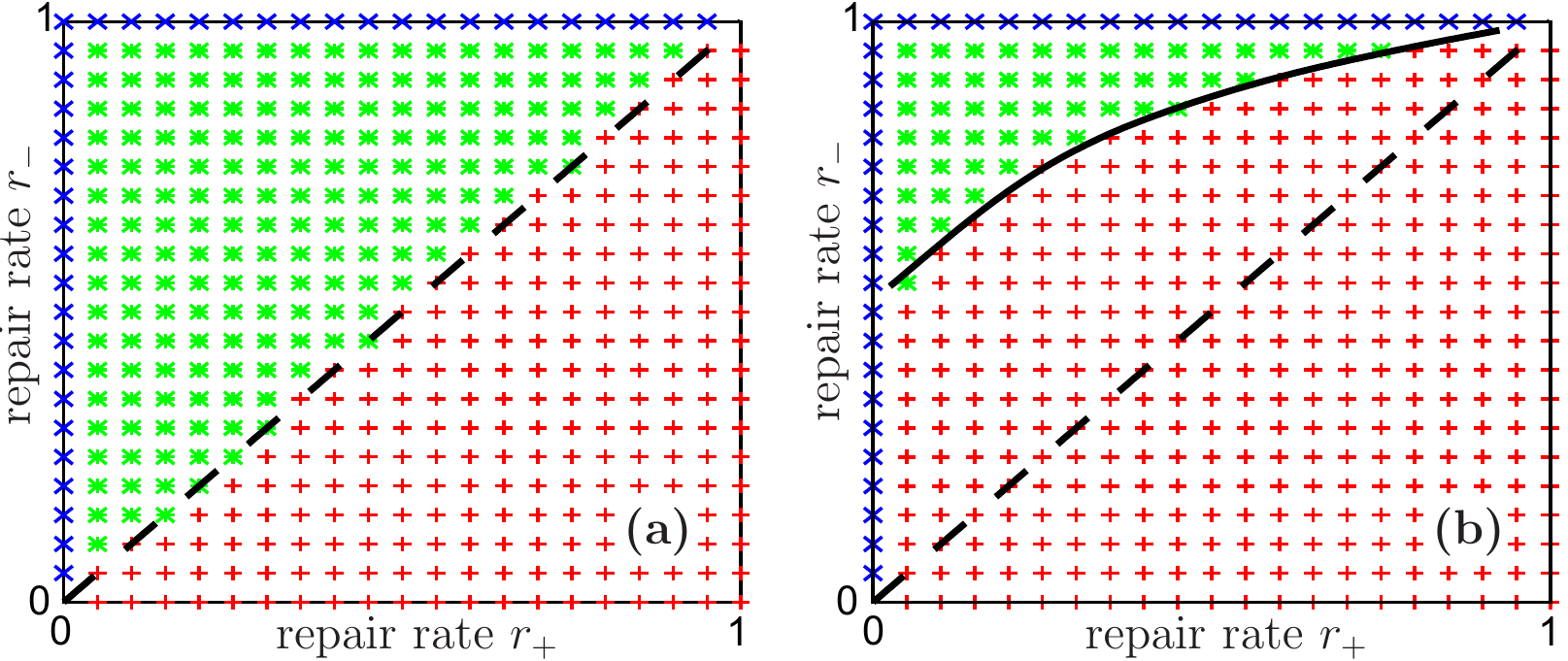}
 \caption{\label{fig:phase_diagrams} \emph{Phase diagrams} for (a) a single random walker on a periodic lattice with varying length $L$ and (b) varying number of random walkers and fixed $L=160$. Dashed lines represent the phase boundary calculated from the FB picture, eq.~\eqref{manyb:cond}. Solid line is a guide to the eye indicating the actual phase boundary. Employing a Gillespie algorithm, the parameter regime has been scanned with a resolution of $\delta r_\pm = 0.05$. Red crosses ({\color{red}+}) indicate a phase with positive velocity, blue crosses ({\color{blue}$\times$})  a phase with negative velocity, and green stars ({\color{green}$\star$}) a phase with current reversal.}
\end{figure}

Equation~\eqref{eq:velocity_single_walker} quantifies the relative frequency of FB pushing and pulling events, and is compared with results obtained from stochastic simulations of the BBM
in fig.~\ref{fig:manyb_singlew}b. As expected there is nice agreement for $r_+ \ll r_- <1$; the analytical result is even exact for  \(r_+=0\). This allows us to use the FB picture, eq.~\eqref{eq:velocity_single_walker}, to determine the precise conditions for velocity reversal:
\begin{equation}
  0<r_+<r_-<1 \, .
  \label{manyb:cond}
\end{equation}
In this regime the velocity becomes zero at a length $L_0={\ln r_+}/{\ln r_-}$, which may be interpreted as a balance of forces exerted on the particle by a broken bond to its front and rear, respectively.  Both of these results agree very well with our numerical results, summarized as a phase diagram in fig.~\ref{fig:phase_diagrams}a.

\section{The burnt-bridge exclusion process}
We now turn to investigate collective properties emerging when there are many random walkers present at a finite density $\rho$. These interact in two ways: local on-site exclusion and non-local interaction through the history of the random walkers, i.e., the sequence of intact and burnt bridges each walker leaves behind in its trail. The effect of on-site interaction alone on the collective properties is well understood. In the ASEP with periodic boundary conditions, the density $\rho$ is homogeneous and there are no correlations in the steady state \cite{Derrida:1998p29316}. Hence the current  follows from a mean-field argument as the product of the difference in forward and backward hopping rates, $\lambda_\pm$, and the probability to find a given site occupied and its neighboring site empty: $j = (\lambda_+ - \lambda_-) \rho (1 - \rho)$.
The sign of the current only depends on the difference between the forward and backward rate, but not on the particle density. The latter regulates the magnitude of the current.

\begin{figure}
\centering
\includegraphics[width=\columnwidth]{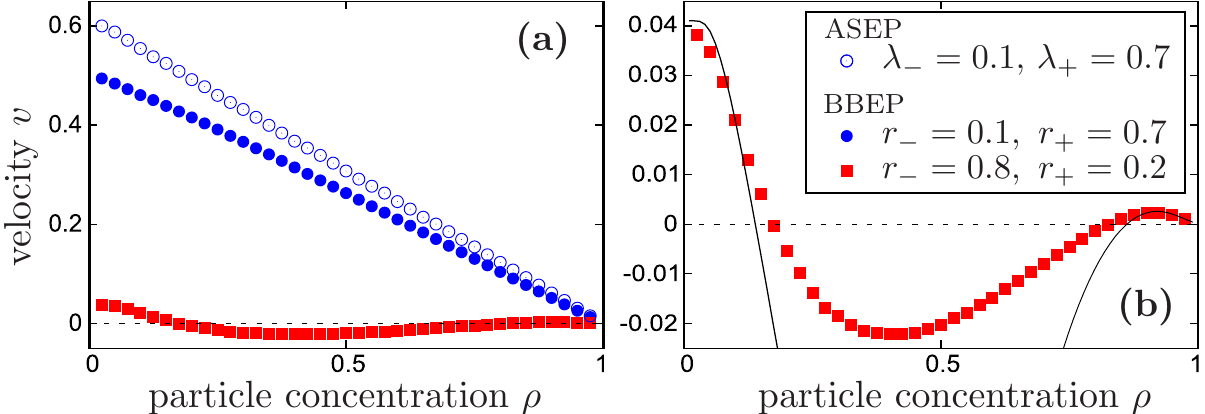} 
\caption{\label{manyp:samples} \emph{Velocity-density relation.} (a) Average particle velocity for the BBEP and ASEP as a function of particle density $\rho$ for a set of kinetic parameters indicated in the graph; $L=40$. Simulations were performed using the Gillespie algorithm. (b) Comparison of simulations with the effective single-particle approximation (solid lines) for $r_-=0.8$ and $r_+=0.2$.}
\end{figure}

In the BBEP with both on-site exclusion and trail-mediated interaction between the random walkers the behavior is much richer. In fig.~\ref{manyp:samples} simulation results for the average velocity $v (\rho):=j/\rho$ as a function of the particle density $\rho$ are shown for both the ASEP and the BBEP. For $r_+ \gg r_-$, we observe a roughly linear velocity-density relation, $v \propto 1-\rho$, closely resembling the relation for the ASEP but with a reduced amplitude. In contrast, for $r_+ \ll r_-$, we find a nonlinear dependence of the velocity on the density $\rho$ strongly deviating from the corresponding ASEP result which would give a negative velocity. In particular, for some parameter ranges, there is \emph{current reversal} at two critical values of the particle density.

We attribute the nonlinear \(v(\rho)\)--relation for \(r_+\ll r_-\) to the history-dependence of the random walks. Every random walker changes the weights of the lattice by either repairing or burning bridges as it traverses the lattice. This affects the dynamics of neighboring particles walking on its trail. What really matters is, however, only the fact that the weights, i.e., state of the bridges, have been changed but not when these changes had occured. This suggests that a walker reacting to the trail of its nearest neighbors might behave quite similar to a single random walker placed on a periodic lattice, always encountering its own trace. In other words, the dynamics of random walkers traveling at a \emph{mean distance}~\(d\) should strongly resemble those of a single particle on a ring of \emph{length}~\(d\), with identical repair rates $r_\pm$. 

\begin{figure}
 \includegraphics[width=\columnwidth]{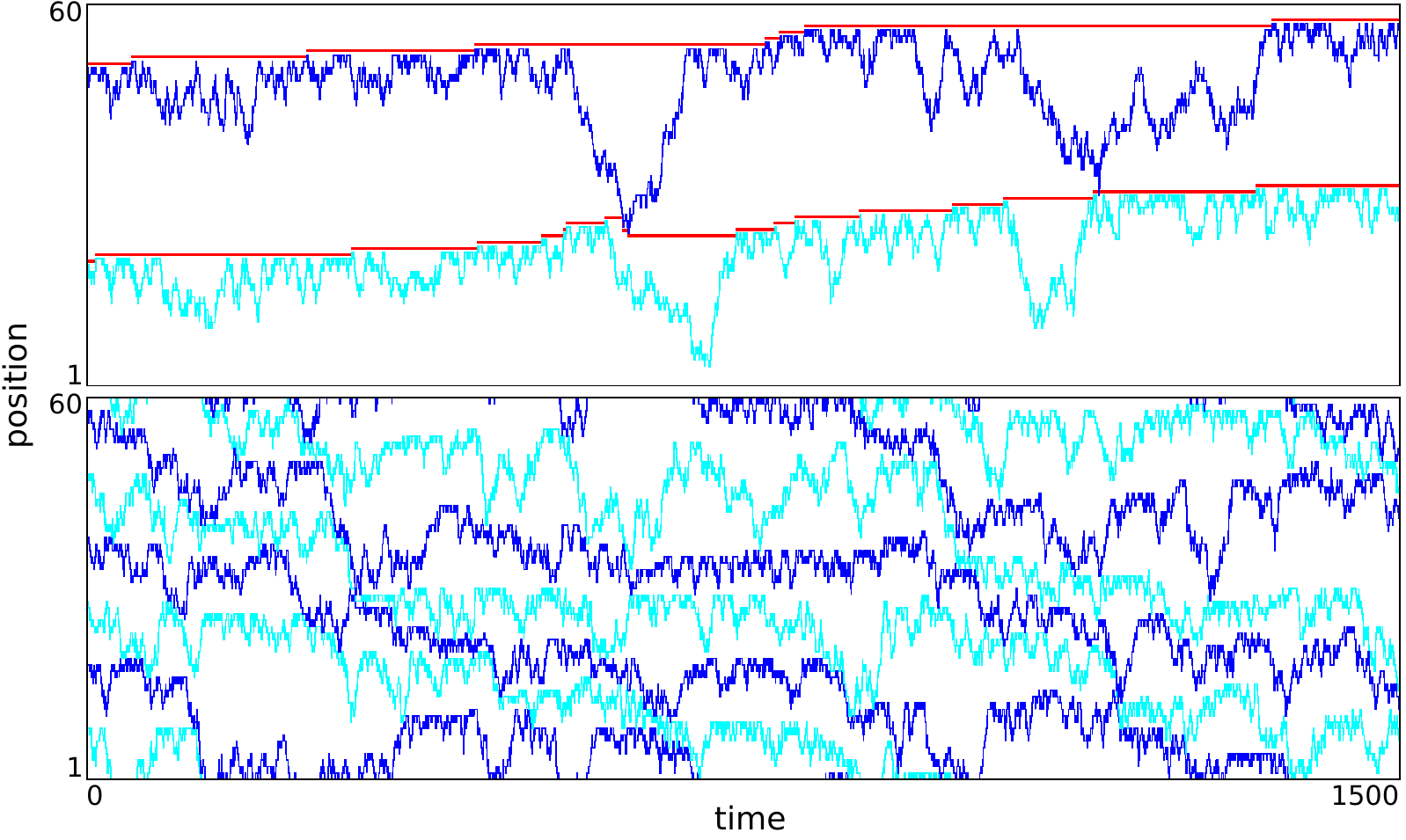}
 \caption{\label{manyp:rw} 
Typical particle trajectories of the BBEP with repair rates \(r_-=0.85\), \(r_+=0.05\), and lattice size  \(L=60\) for two (top) and six (bottom) particles. The trajectories are alternately drawn in blue (dark grey) and turquoise (light gray). For better visualization of the trajectories the bridges' states are not shown. Front barriers are only shown in the top but not in the bottom figure.}
\end{figure}

To illustrate this argument, we compare trajectories of random walkers for a few and many particles in fig.~\ref{manyp:rw}. With each particle we can, at any point in time, associate an unambiguously defined FB. With two random walkers on the lattice, direct encounters are observed only rarely. Most of the time, both are continuously pushing their own FB forward without ever taking notice of the other walker's presence. Only in one rare event, the upper walker reaches the FB of its companion and pulls it back by one lattice unit. This happens only occasionally, and for large times, the average velocity is positive (in the forward direction). In contrast, if there are many particles on the lattice pulling events become more frequent simply because there is a shorter mean distance between the walkers. As a consequence, for the parameter values chosen in fig.~\ref{manyp:rw}, the average particle velocity is reversed since pulling events outnumber pushing events of FBs.

Upon replacing the lattice length $L$ by the mean particle spacing \(d=1/\rho\) in eq.~\eqref{eq:velocity_single_walker} one obtains an effective single-particle approximation for the many-body system. As illustrated in fig.~\ref{manyp:samples}b, it gives a reasonable description of the velocity-density relation at small densities but becomes quantitatively wrong with increasing density. This is to be expected since correlations due to on-site exclusion are not accounted for in the FB picture. The results obtained from the FB picture for low densities can be generalized to very high densities close to $\rho=1$ upon employing \emph{particle-hole symmetry}, which implies $\rho  v(\rho)=(1-\rho) v(1-\rho)$. 

The single-particle front barrier picture qualitatively explains the observed current reversal in the many-particle BBEP. It identifies competition between pushing and pulling events of the FBs as the main mechanism. This trail-mediated mechanism is distinct from other current reversal mechanisms observed in Brownian ratchet models \cite{DERENYI:1995p30245}. There an interplay of the local ratcheting mechanism and hard-core repulsion between the particles leads to a reversal of the net current with increasing particle density. While in the latter case, excluded volume effects are driving current reversal, they actually weaken it in the BBEP.
As can be inferred from fig.~\ref{manyp:samples}b the actual reduction in the average particle current with density is less than the FB picture would predict. Crowding effects asymmetrically affect the frequency of pulling and pushing events, to the disadvantage of pulling events. 

To map out the phase diagram we performed extensive stochastic simulations for the BBEP and measured velocity-density relations for a broad range of repair rates; see fig.~\ref{fig:phase_diagrams}b. Indeed, there is a broad parameter regime, where current reversal is observed; crowding effects, however, narrow this regime as compared to the results obtained from the effective single-particle picture which considers trail-mediated effects only. 

\section{Summary}
The burnt-bridge exclusion process has features which are genuinely distinct from the Markovian analog asymmetric exclusion process. The most prominent is current reversal as a function of density. This phenomenon is absent for the asymmetric exclusion process and is attributed to the non-local interaction between the walkers through their trails. Those interactions are caused by a strong coupling between the dynamics of the particles and the lattice which lead to memory effects acting opposite to the broken bonds' asymmetric transmissibility. The observed phenomenon is robust to model variations \cite{LongPaper}. We suppose that exclusion processes for the broader class of random walks on lattices with weighted bonds or sites shows an even richer dynamics with a plethora of new phenomena, e.g. by considering non-Markovian analogs of generalized exclusion processes \cite{pff,rff,gkr}. It will be also interesting to test our predictions using artificial systems such as molecular spiders and molecular robots.

\acknowledgments
We are grateful to Louis Reese for useful discussions. The present research was supported by the German Excellence Initiative via the program ``Nanosystems Initiative Munich", the Deutsche Forschungsgemeinschaft via SFB TR12  (EF), the Welch Foundation (Grant No. C-1559), and the U.S. National Institute of Health (Grant No. R01GM094489) (AKB).


\begin{thebibliography}{0}

\bibitem{Julicher:1997p28955}
 \Name{J\"ulicher F., Ajdari A. \and Prost J.}
 \REVIEW{Rev. Mod. Phys.}{69}{1997}{1269}.

\bibitem{reimann-2002-361}
 \Name{Reimann P.}
 \REVIEW{Phys. Rep.}{361}{2002}{57}.

\bibitem{Haenggi:2009p29024}
 \Name{Haenggi P. \and Marchesoni F.}
 \REVIEW{Rev. Mod. Phys.}{81}{2009}{387}.

\bibitem{Kay:2007p28153}
 \Name{Kay E. R., Leigh D. A. \and Zerbetto F.}
 \REVIEW{Angew. Chem. Int. Ed.}{46}{2007}{72}.

\bibitem{Saffarian:2004p28152}
 \Name{Saffarian S. \etal}
 \REVIEW{Science}{306}{2004}{108}.

\bibitem{Saffarian:2006p28147}
 \Name{Saffarian S. \etal}
 \REVIEW{Phys. Rev. E}{73}{2006}{041909}.

\bibitem{Mai:2001p27900}
 \Name{Mai J., Sokolov I. M. \and Blumen A.}
 \REVIEW{Phys. Rev. E}{64}{2001}{011102}.

\bibitem{Klapstein:2000}
 \Name{Klapstein K. D. \and Bruinsma R.}
 \REVIEW{J. Bio. Chem.}{275}{2000}{16073}.

\bibitem{Lakhanpal:2007}
\Name{Lakhanpal A. \and Chou T.}
 \REVIEW{Phys. Rev. Lett.}{99}{2007}{248302}.


\bibitem{Omabegho:2009p67}
 \Name{Omabegho T., Sha R. \and Seeman N. C.}
 \REVIEW{Science}{324}{2009}{67}
 
\bibitem{Pei:2006p27865}
 \Name{Pei R. \etal}
 \REVIEW{J. Am. Chem. Soc.}{128}{2006}{12693}.

\bibitem{Samii:2010p021106}
 \Name{Samii L. \etal}
 \REVIEW{Phys. Rev. E}{81}{2010}{021106}
 
\bibitem{Semenov:2011p021117}
 \Name{Semenov O., Olah M. J. \and Stefanovic D.}
 \REVIEW{Phys. Rev. E}{83}{2011}{021117}

\bibitem{Lund:2010p28160}
 \Name{Lund K. \etal}
 \REVIEW{Nature}{465}{2010}{206}.

\bibitem{Madras_Slade:book}
 \Name{Madras N. \and Slade G.}
 \Book{The Self-Avoiding Walk} \Publ{Birkh{\"a}user, Boston} \Year{1996}.

\bibitem{DAVIS:1990p28173}
 \Name{Davis B.}
 \REVIEW{Probab. Theory Rel.}{84}{1990}{203}.

\bibitem{Antal:2005p28234}
 \Name{Antal T. \and Redner S.}
 \REVIEW{J. Phys. A}{38}{2005}{2555}.

\bibitem{Derrida:1998p29316}
 For a review see e.g.: \Name{Derrida B.}
 \REVIEW{Phys. Rep.}{301}{1998}{65}.

\bibitem{DERENYI:1995p30245}
 \Name{Derenyi I. \and Vicsek T.}
 \REVIEW{Phys. Rev. Lett.}{75}{1995}{374}.

\bibitem{Antal:2005p27853}
 \Name{Antal T. \and Krapivsky P. L.}
 \REVIEW{Phys. Rev. E}{72}{2005}{046104}.

\bibitem{Morozov:2007p9187}
 \Name{Morozov A. Y., Pronina E., Kolomeisky A. B. \and Artyomov M. N.}
 \REVIEW{Phys. Rev. E}{75}{2007}{031910}.

\bibitem{Artyomov:2008p27887}
 \Name{Artyomov M. N., Morozov A. Y. \and Kolomeisky A. B.}
 \REVIEW{Phys. Rev. E}{77}{2008}{040901}.

\bibitem{Artyomov:2010p27938}
 \Name{Artyomov M. N., Morozov A. Y. \and Kolomeisky A. B.}
 \REVIEW{Condens. Matter Phys.}{13}{2010}{23801}.


\bibitem{LongPaper}
 \Name{Schulz J. H. P., Kolomeisky A. B. \and Frey E.}
 \Page{in preperation}.

\bibitem{pff}
 \Name{Parmeggiani A., Franosch T. \and Frey E.}
 \REVIEW{Phys. Rev. Lett.}{90}{2003}{086601}.
 
 \bibitem{rff}
 \Name{Reichenbach T., Franosch T. \and Frey E.}
 \REVIEW{Phys. Rev. Lett.}{97}{2006}{050603}.

\bibitem{gkr}
 \Name{Gabel A., Krapivsky P. L. \and Redner S.}
 \REVIEW{Phys. Rev. Lett.}{105}{2010}{210603}.

\end{thebibliography}
\end{document}